\documentclass[journal=ancac3, manuscript=article, layout=twocolumn]{achemso}
\usepackage{latexsym, amsmath, amssymb}
\usepackage[english]{babel}
\usepackage[T1]{fontenc}
\usepackage[utf8]{inputenc}
\usepackage{bm}
\usepackage{color}
\usepackage{soul}

\newcommand{\Fig}[2]{Figure \ref{#1}#2}

\newcommand{\pressure}[2]{$#1 \cdot 10^{-#2}$ mbar}
\newcommand{\supp}{Supporting Information}

\title{A Molecular Approach for Engineering Interfacial Interactions in Magnetic-Topological Insulator Heterostructures}

\author{Marc G. Cuxart}
\email{marc.gonzalez-cuxart@tum.de}
\affiliation{Catalan Institute of Nanoscience and Nanotechnology (ICN2). CSIC and BIST, Campus UAB, 08193 Barcelona, Spain.}

\author{Miguel Angel Valbuena}
\affiliation{Catalan Institute of Nanoscience and Nanotechnology (ICN2). CSIC and BIST, Campus UAB, 08193 Barcelona, Spain.}
\author{Roberto Robles}
\affiliation{Centro de F\'{i}sica de Materiales CFM/MPC (CSIC-UPV/EHU), 20018 Donostia-San Sebasti\'an, Spain.}

\author{César Moreno}
\affiliation{Catalan Institute of Nanoscience and Nanotechnology (ICN2). CSIC and BIST, Campus UAB, 08193 Barcelona, Spain.}

\author{Frédéric Bonell}
\affiliation{Catalan Institute of Nanoscience and Nanotechnology (ICN2). CSIC and BIST, Campus UAB, 08193 Barcelona, Spain.}

\author{Guillaume Sauthier}
\affiliation{Catalan Institute of Nanoscience and Nanotechnology (ICN2). CSIC and BIST, Campus UAB, 08193 Barcelona, Spain.}

\author{Inhar Imaz}
\affiliation{Catalan Institute of Nanoscience and Nanotechnology (ICN2). CSIC and BIST, Campus UAB, 08193 Barcelona, Spain.}

\author{Heng Xu}
\affiliation{Catalan Institute of Nanoscience and Nanotechnology (ICN2). CSIC and BIST, Campus UAB, 08193 Barcelona, Spain.}

\author{Corneliu Nistor}
\affiliation{Department of Materials, ETH Zurich, H\"{o}nggerbergring 64, CH-8093 Zurich, Switzerland.}

\author{Alessandro Barla}
\affiliation{Istituto di Struttura della Materia (ISM), Consiglio Nazionale delle Ricerche (CNR), I-34149 Trieste, Italy.}

\author{Pierluigi Gargiani}
\affiliation{ALBA Synchrotron Light Source, Carretera BP 1413km 3.3, E-08290 Cerdanyola del Vall\` es, Spain.}

\author{Manuel Valvidares}
\affiliation{ALBA Synchrotron Light Source, Carretera BP 1413km 3.3, E-08290 Cerdanyola del Vall\` es, Spain.}

\author{Daniel Maspoch}
\affiliation{Catalan Institute of Nanoscience and Nanotechnology (ICN2). CSIC and BIST, Campus UAB, 08193 Barcelona, Spain.}
\alsoaffiliation{Institució Catalana de Recerca i Estudis Avançats (ICREA), Barcelona 08070, Spain.}

\author{Pietro Gambardella}
\affiliation{Department of Materials, ETH Zurich, H\"{o}nggerbergring 64, CH-8093 Zurich, Switzerland.}

\author{Sergio O. Valenzuela}
\email{SOV@icrea.cat}
\affiliation{Catalan Institute of Nanoscience and Nanotechnology (ICN2). CSIC and BIST, Campus UAB, 08193 Barcelona, Spain.}
\alsoaffiliation{Institució Catalana de Recerca i Estudis Avançats (ICREA), Barcelona 08070, Spain.}

\author{Aitor Mugarza}
\email{aitor.mugarza@icn2.cat}
\affiliation{Catalan Institute of Nanoscience and Nanotechnology (ICN2). CSIC and BIST, Campus UAB, 08193 Barcelona, Spain.}
\alsoaffiliation{Institució Catalana de Recerca i Estudis Avançats (ICREA), Barcelona 08070, Spain.}


\begin{document}

\begin{abstract}
 \textbf{
Controlling interfacial interactions in magnetic/topological insulator heterostructures is a major challenge for the emergence of novel spin-dependent electronic phenomena. As for any rational design of heterostructures that rely on proximity effects, one should ideally retain the overall properties of each component while tuning interactions at the interface. However, in most inorganic interfaces interactions are too strong, consequently perturbing, and even quenching, both the magnetic moment and the topological surface states at each side of the interface. Here we show that these properties can be preserved by using ligand chemistry to tune the interaction of magnetic ions with the surface states. By depositing Co-based porphyrin and phthalocyanine monolayers on the surface of Bi$_2$Te$_3$ thin films, robust interfaces are formed that preserve undoped topological surface states as well as the pristine magnetic moment of the divalent Co ions. The selected ligands allow us to tune the interfacial hybridization within this weak interaction regime. These results, which are in stark contrast with the observed suppression of the surface state at the first quintuple layer of Bi$_2$Se$_3$ induced by the interaction with Co phthalocyanines, demonstrate the capability of  planar metal-organic molecules to span interactions from the strong to the weak limit.}

\end{abstract}

\section{Introduction}

Interfacing topological insulators (TI) with magnetic materials can give rise to interesting phenomena that result from breaking time reversal symmetry, such as band-gap opening in the topological surface states (TSS) and the emergence of the quantum anomalous Hall effect.\cite{Hasan2010a,Chen2010,Chang2013} It can also enable the interconversion of spin and charge currents.\cite{Rojas-Sanchez2016} The observation of these phenomena, however, demands a fine control over the Fermi level position,\cite{Winnerlein2017,Kondou2016} and the interfacial hybridization.\cite{MenShov2013}

A way of engineering these properties in a predictive and controlled manner is by tuning interfacial interactions without substantially perturbing the pristine properties of each component, such that the magnetic and topological materials can be designed independently. Unfortunately, this is not the case when the TI is interfaced directly with metallic ferromagnets.\cite{Li2012, Spataru2014, DeJong2015, Zhang2016b, Walsh2017} The strong interaction leads to undesired effects such as magnetically dead or ill-defined layers induced by intermixing,\cite{Walsh2017} or the suppression of the TSS.\cite{Rojas-Sanchez2016} At the single impurity level the tendency for intermixing expresses as strong lattice relaxations around the impurity atom, leading to multiple configurations that hamper the reliable control of the doping of the TSS and of the magnetic moment and anisotropy of the impurities.\cite{Wray2010, Honolka2012, Scholz2012,Schlenk2013, Eelbo2014, Martinez-Velarte2017} The magnetic exchange required for long-range magnetic ordering is also hard to control. The interatomic distance and doping level, which are tightly entangled via the impurity concentration, can alter the size and signature of the interaction by the interplay between the different exchange mechanisms.\cite{Russmann2018} 

An effective way of regulating the parameters that govern all of the above factors is to coordinate the single magnetic impurities with organic ligands. Using this approach, the interaction between the magnetic ion and the TSS can be finely tuned with ligands that place the magnetic ion at a selected distance from the surface.\cite{Zhao2005} Furthermore, molecular self-assembly interconnects magnetic impurities and arrange them with regular interatomic spacing, facilitating control over Ruderman–Kittel–Kasuya– Yoshida (RKKY) interactions, and promoting ligand-mediated interionic exchange. Indeed, ligand-mediated long-range magnetism in a metal-organic coordination network in contact with a TI surface has been predicted to break time-reversal symmetry and open a band-gap at the Dirac point of the TSS.\cite{Otrokov2015} The interaction between the organic ligand and the TSS can additionally be exploited to tune the spin-charge interconversion.\cite{Nakayama2018} Ultimately, metal-organic frameworks with coexisting magnetic and electric orderings\cite{Xu2011a} might be used to simultaneous induce magnetic proximity and electrostatic gating of TI interfaces, or as tunable buffer layers for controlling the spin-charge conversion efficiency to an adjacent ferromagnetic metal.

The degree of tunability of charge and spin interactions offered by metal-organic layers has already been demonstrated on metallic surfaces.\cite{Cinchetti2017, Zhao2005, Iancu2006, Cinchetti2009, Auwarter2010,LodiRizzini2011, Mugarza2011,Mugarza2012, Raman2013, Gruber2015} The few systematic studies on TIs suggest a similar tunability, driven by either the choice of the metal ion \cite{Bathon2015} or the ligand.\cite{Jakobs2015} Surprisingly, despite the considerably larger distance from the surface, as compared to bare impurities, interactions between the molecular metal ions and the TSS can still be strong enough to lead to doping levels comparable to those observed for bare impurities,\cite{Sk2018,Caputo2016} form hybrid interface states \cite{Song2014,Sessi2014,Hewitt2017} and, more dramatically suppress the TSS from the first quintuple layer (QL), as recently found for cobalt phthalocyanine (CoPc) on Bi$_2$Se$_3$.\cite{Caputo2016} A follow-up study concluded that non-perturbative interactions can only be achieved with molecules with non-planar coordination geometry, where the metal ion is protected by the ligand core from surface-induced perturbations. In this configuration, however, magnetic interactions are limited to dipolar and, as a consequence, relatively weak.\cite{Nam2018}

In order to validate the approach of planar coordination geometries, we provide here a systematic investigation of the interfacial properties of a cobalt tetrakis (4-bromophenyl) porphyrin (CoTBrPP) monolayer deposited on the (0001) surface of a twin-free Bi$_2$Te$_3$ thin film.\cite{Bonell2017} By combining scanning tunneling microscopy and spectroscopy (STM/STS), angle-resolved photoelectron spectroscopy (ARPES), and x-ray absorption and magnetic circular dichroism (XAS/XMCD), we are able to probe the TSS band dispersion and both the molecular orbital structure and its magnetic moment. We find that doping effects are negligible, and that the magnetic moment of Co is neither quenched nor Kondo screened. Yet, our STS analysis, supported by Density Functional Theory (DFT) calculations, reveal that the spin carrying Co $d_{z^2}$ orbital undergoes an appreciable hybridization with substrate electronic states, as required for magnetic exchange interactions. Our results show that non-perturbative interfaces can also be attained by using metal-organic molecules that expose the magnetic ion directly to the surface, if the ligand and substrate are appropriately chosen. By replacing the porphyrin ligand with the more planar phthalocyanine, we further demonstrate that the position of the Fermi level and the interfacial hybridization can be tuned within this interaction regime by selecting different ligands.

\section{Results and discussion}
\subsection{Structural characterization}

\begin{figure*}
	\includegraphics{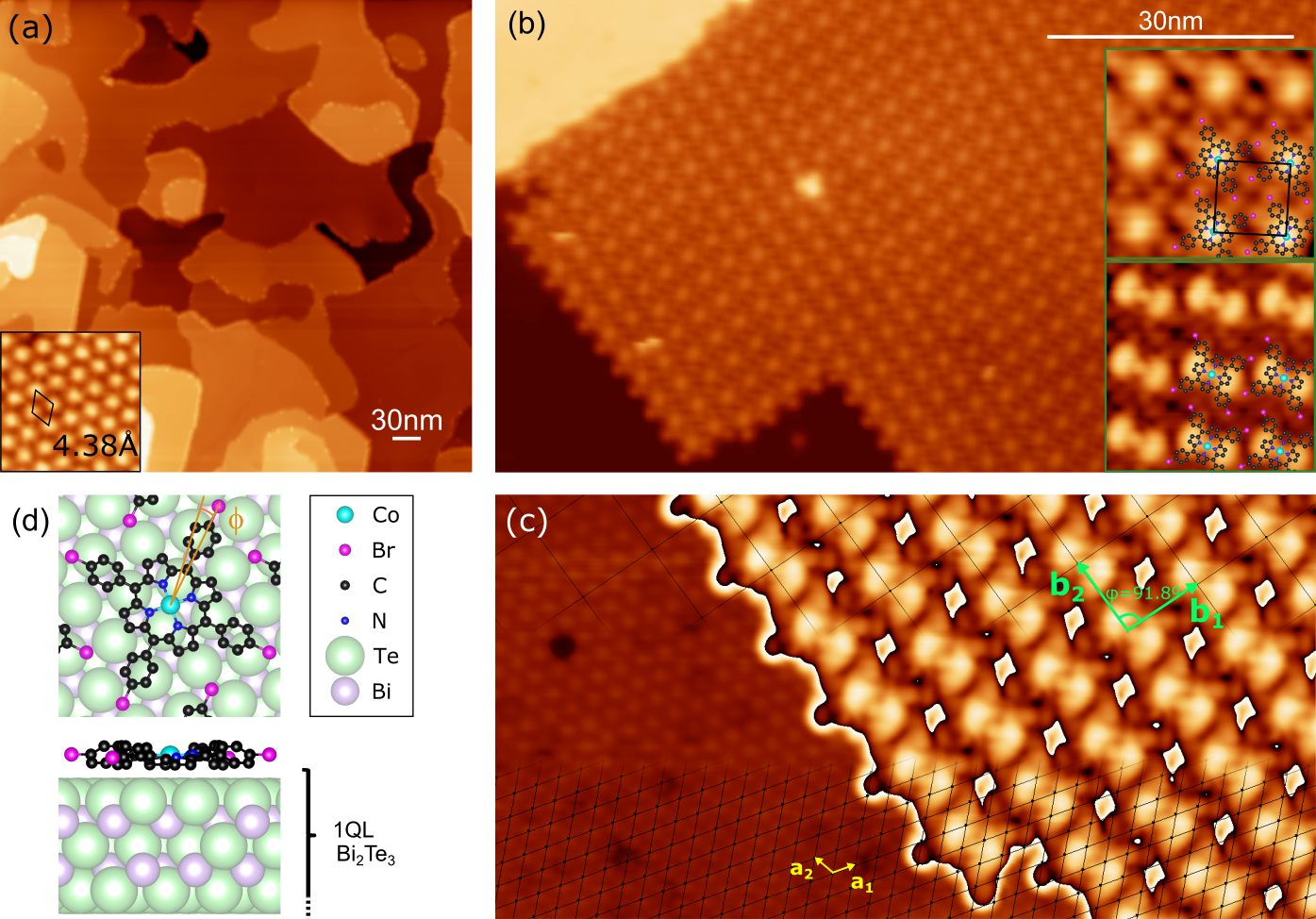}
	\caption{STM topographic images of the Bi$_2$Te$_3$ substrate (a) after the Te cap removal ($V_{bias}= 1.4 \ V$, $I_t= 20 \ pA$). Inset: atomically resolved flat terrace with its unit cell ($V_{bias}= 0.4 \ V$, $I_t= 540 \ pA$); and (b) after the deposition of CoTBrPP molecules ($V_{bias}=1.4$ V, $I_t=10$ pA). Insets are close-up images of the molecular lattice obtained at $V_{bias}=1.6$ V and $V_{bias}=-1.2$ V respectively. (c) STM image of a CoTBrPP island on Bi$_2$Te$_3$ with combined atomic and intramolecular resolution ($V_{bias}=-1.8$ V, $I_t=10$ pA). The surface and molecular lattices, and respective unit cells are overlaid. (d) Top and side view of the minimum energy adsorption configuration of CoTBrPP on Bi$_2$Te$_3$(0001), as obtained by DFT.}
	\label{STM}
\end{figure*}

The structural characterization of the Bi$_2$Te$_3$ surface and the molecular overlayer carried out by STM is summarized in \Fig{STM}. After the thermal desorption of a protective Te layer (see Methods), the surface of the Bi$_2$Te$_3$ thin films exhibit atomically flat terraces exposing the outermost Te atoms arranged in a hexagonal lattice (see \Fig{STM}{a}). After submonolayer deposition of CoTBrPP, self-assembled molecular islands grow from the step edges. Extended regular islands separated by large areas of clean surface are formed even at low coverage, revealing a high diffusion rate of the molecules, a first indication of weak interaction with the substrate. The molecular appearance changes as we probe different Co-derived orbitals by varying the bias voltage (inset in \Fig{STM}{b}), resulting in bright circular centers at positive bias, and dumbbell-like structures oriented parallel to one of the pyrrole-pyrrole axial directions at negative bias. 

A complete description of the molecular lattice can be obtained from the  atomically resolved STM image shown in \Fig{STM}{c}. By extrapolating the  Bi$_2$Te$_3$ surface lattice to the molecular island or, conversely, the molecular lattice to the Bi$_2$Te$_3$ surface, we derive the stacking relation and relative angle of the two lattices, as well as the azimuthal angle of the molecular axis relative to the surface lattice. The molecular lattice has a commensurate stacking with the surface, with Co ions sitting on Te bridge sites, in agreement with the most stable adsorption configuration found in DFT calculations (see \Fig{STM}{d}). The respective experimental and theoretical azimuthal rotation of the phenyl axis, of $\phi=9.0 \pm2.8^{\circ}$ and $\phi=9.8^{\circ}$, are also in close agreement. The molecular lattice vectors ($\bm{b_}1$,$\bm{b_2}$), with a modulus of $|\bm{b_1}|=|\bm{b_2}|=16.6 \pm 0.7$ \AA, form an angle of $\varphi = 91.9 \pm 0.8^{\circ}$. The slight deviation from the otherwise favourable squared lattice is a consequence of keeping registry with the underlying hexagonal lattice of the surface, with a relation of $\bm{b_1}=4\bm{a_1}+\bm{a_2}$ and $\bm{b_2}=\bm{a_1}+4\bm{a_2}$.

A previous systematic study of the molecular self-assembly of different metal phthalocyanines on Bi$_2$Se$_3$ nicely illustrates the trend from a frustrated assembly for the strongly interacting MnPc, to a hexagonal arrangement mimicking the underlying surface lattice for the intermediate case of CoPc and, finally, to an incommensurate square arrangement for the weakly interacting CuPc.\cite{Bathon2015} The commensurate, quasi-square arrangement that is observed for CoTBrPP on Bi$_2$Te$_3$ lies somewhere between the intermediate and weak interaction cases of CoPc and CuPc on Bi$_2$Se$_3$.

\subsection{Molecular orbitals}

\begin{figure*}
	\includegraphics{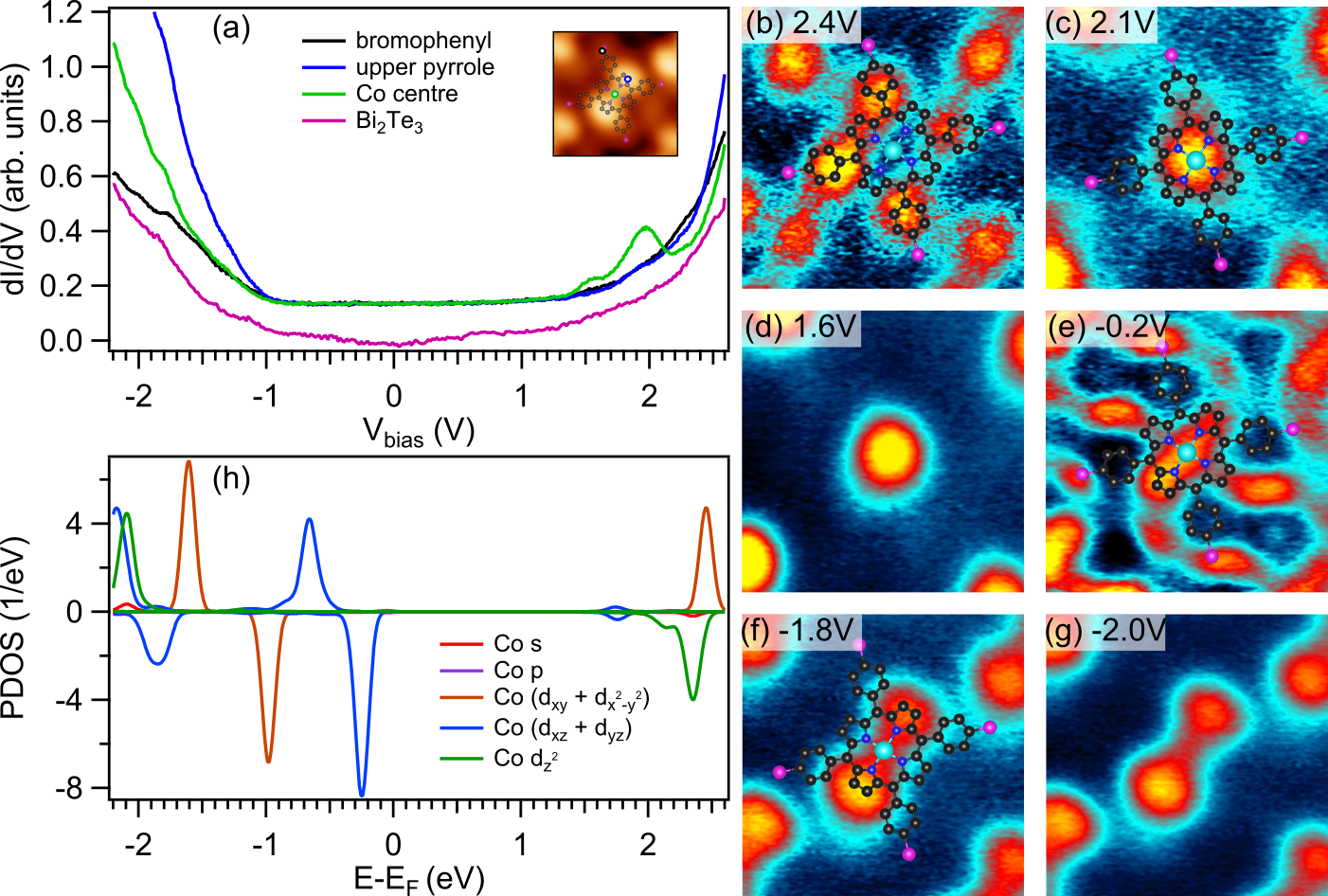}
	\caption{(a) $dI/dV$ spectra acquired  on a CoTBrPP molecule within an island at the points depicted in the topographic image at the inset, and on the bare Bi$_2$Te$_3$ surface.  Spectra acquired on the molecule are vertically shifted for visualizing the finite $dI/dV$ signal of the bare surface around E$_F$ (constant-height mode, $V_{bias}=-1.4$ V, $I_t=10$ pA, image size $2.5 \times 2.5$ nm$^2$). (b) - (g) Constant height $dI/dV$ maps of the same molecule. (h) PDOS of the Co orbitals calculated by DFT.}
	\label{STS}
\end{figure*}

Next, the molecular electronic structure is investigated by performing local STS measurements on single CoTBrPP molecules within the islands. Spectra acquired at representative sites of the molecule are represented in \Fig{STS}{a}, together with that acquired on the bare substrates (see \supp\ for further details on the spectra on the bare substrate). Resonant features can be assigned to molecular orbitals by comparing their corresponding maps (\Fig{STS}{b}-g) with the projected density of states (PDOS) on different atomic orbitals  (\Fig{STS}{h}).

The first noticeable characteristic of the $dI/dV$ spectra is a large gap of $\sim 2.4$ V defined by the onsets of the highest occupied (HOMO) and lowest unoccupied (LUMO) molecular orbitals at $-1.0$ V and $1.4$ V. In contrast to the smaller gaps of $\sim 1$ eV observed for Co porphyrins on noble metals,\cite{Auwarter2010,Iancu2006b} the proximity of the measured value to the 2.5 V calculated for the free-standing molecule \cite{Liao2002a} indicates that molecular states are effectively protected from strong surface hybridization and screening effects. 

The HOMO onset is most pronounced in the spectrum acquired on the pyrrole sites (the dumbbell protrusion pair seen in the images acquired at negative bias, see inset in \Fig{STS}{a}). Constant height conductance maps acquired at energies below this onset replicate the dumbbell feature, which originates from hybrid Co $d_{xz,yz}$  and pyrrole $p$ states.\cite{Auwarter2010} The two-fold symmetry of the feature is related to the saddle distortion of adsorbed CoTBrPP, which reduces the molecular symmetry from four to two-fold and consequently splits the $d_{xz}$ and $d_{yz}$ orbitals.

The LUMO is the broad structured resonance in the spectrum acquired on the Co ion. Indeed, conductance maps around the LUMO onset at $+1.6$ V reveal a spherical protrusion localized on the Co ion, suggesting the presence of a $d_{z^2}$ orbital in this energy region. The spherical feature persists at higher energies up to the end of the resonance at $\sim2.1$ V, above which the spectral weight shifts to the peripheral phenyls. 

For an unequivocal identification of Co $d$ orbital signatures in the spectral features, we compare the latter with the spin-resolved PDOS of the atomic orbitals. \Fig{STS}h depicts the PDOS of the Co $d$ orbitals, which define the molecular gap and magnetic moment (more extended data with projection to ligand atoms is provided in the \supp). A gap of $\sim 2$ eV, similar to the experimental one, is obtained by using a U-J=3 eV term, a value comparable to that used in other weakly interacting Co porphyrin compounds.\cite{Leung2010,Hermanns2013} The orbital sequence suggested by the conductance maps is also confirmed, since the HOMO and LUMO are respectively of $d_{xz,yz}$ and $d_{z^2}$ character in the PDOS. Interestingly, the unoccupied $d_{z^2}$ orbital also appears as a broad double-peak resonance, emulating the observed experimental feature at positive bias, and indicating that hybridization with the underlying Te states is not negligible.

%The effect of the saddle distortion is reflected in a large splitting of the $d_{xz}$ and $d_{yz}$ orbitals of around 1.6 eV.

The structural and spectroscopic analysis of the molecular films reveal a finite albeit moderate interaction of the Co ions with the substrate. The effect of this interaction on the magnetic moment of the Co ion and the TSS is discussed in the next Section.

\subsection{Molecular magnetic moment}

\begin{figure}
	\includegraphics{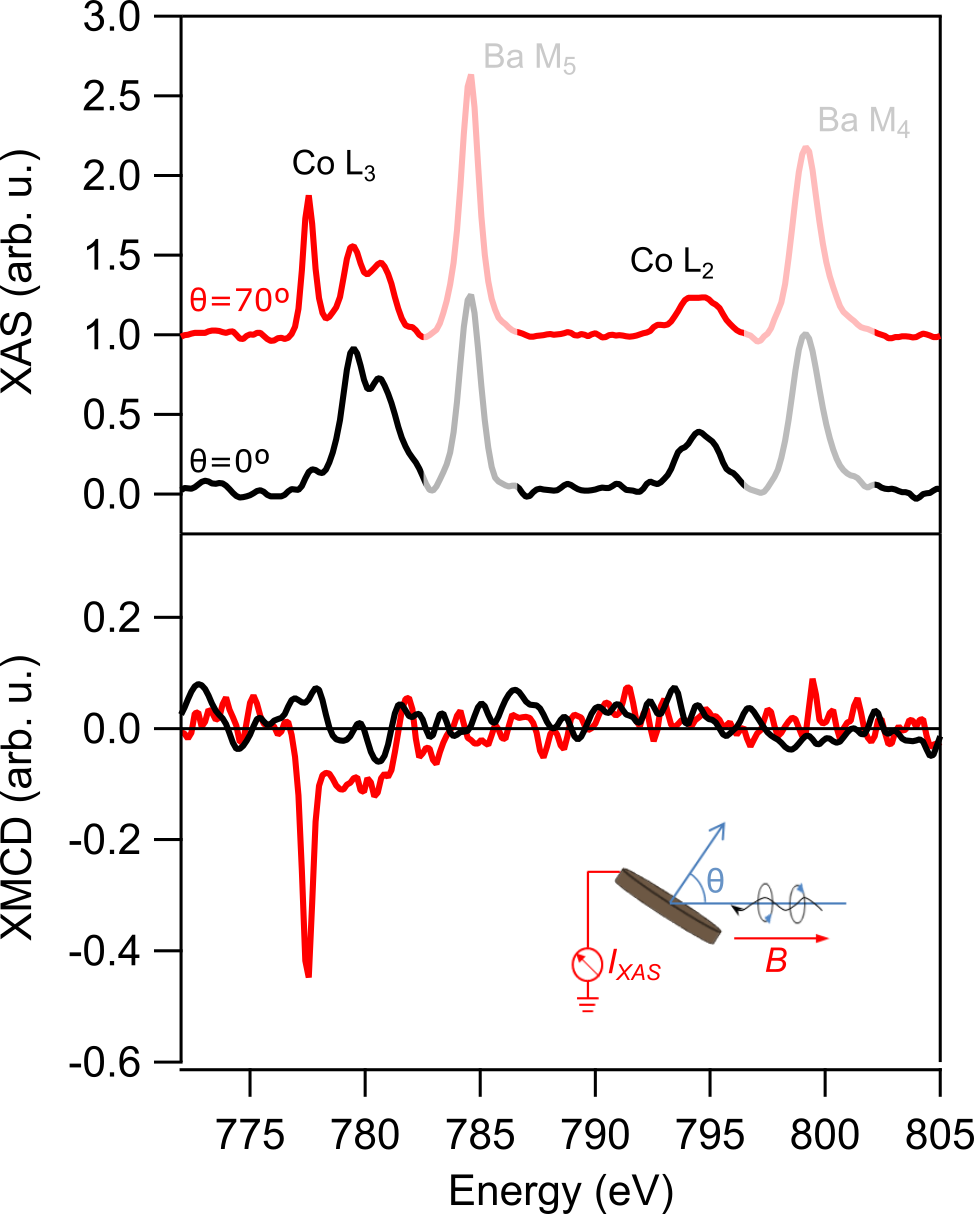}
	\caption{(a) Average XAS and XMCD measured on  a nearly saturated monolayer of CoTBrPP on Bi$_2$Te$_3$ at the Co $L_{2, 3}$ edge at normal ($\theta=0^{\circ}$) and grazing ($\theta=70^{\circ}$) incidence, with a magnetic field of 6 T and at a temperature of 7.5 K. Features at 784 and 799 eV are attributed to the Ba $M_{4,5}$ edge from the BaF$_2$ substrate.}
	\label{XMCD}
\end{figure}

The magnetic moment of the Co ion can be directly probed by XMCD. \Fig{XMCD}{} shows XAS and XMCD spectra acquired for a nearly completed CoTBrPP monolayer at the energy range of the Co $L_{2,3}$ edge. The spectra are acquired at normal and grazing incidence, which probe the out-of-plane and in-plane projection of the magnetic moment of the Co 3$d$ orbitals respectively. Since photoelectrons can still be emitted across the Bi$_2$Te$_3$ film, a Ba $M_{4,5}$ doublet arising from the underlying BaF$_2$ substrate is also visible. The overlap between the Ba $M_{4,5}$ and Co $L_{2,3}$ edges components is however minimum and does not impede a qualitative analysis of the Co $L_{2,3}$ lineshape. 

At grazing incidence, the Co absorption spectrum is characterized by an $L_3$ multiplet with a sharp intense peak on the low energy side followed by a doublet, and a broad $L_2$ peak. At normal incidence, the low energy  peak of the $L_3$ component is practically absent, whereas the intensity of the $L_3$ doublet and $L_2$ component increase. The overall lineshape and its angular dependence closely resemble those of a Co$^{2+}$ ion in a square planar ligand field with weak interactions with the substrate, such as in CoPc multilayers \cite{Stepanow2011} and CoOEP on graphene.\cite{Hermanns2013} In analogy with these systems, the x-ray absorption spectra thus indicate that the electronic configuration of the Co ions is characterized by an unpaired electron in the $a_{1g}(d_{z^2})$ orbital that gives rise to a S=1/2 ground state. Spin-orbit coupling additionally leads to mixing of electronic configurations with $2A_{1g}$ and $2E_g$ symmetry, which results in charge transfer from the $e_g$ to the $a_{1g}$ orbital and a finite in-plane orbital moment.\cite{Stepanow2011}

Consistently with the S=1/2 ground state inferred from the XAS lineshape, the XMCD spectra measured at grazing incidence reveal the presence of a finite magnetic moment residing on the Co ions. The dichroic signal is dominated by a sharp peak at the low-energy $L_3$ component, which corresponds to transitions to the half-filled $d_{z^2}$ orbital. The signal vanishes within the noise level at normal incidence due to the strong anisotropy of the $d_{z^2}$ spin density, which results in a pronounced angle-dependent dipolar term.\cite{Stepanow2010} The absence of XMCD spectral intensity at the $L_2$ edge further indicates that the Co ions have a relatively strong orbital magnetic moment,\cite{Gambardella2002c,Bartolome2010} as anticipated from the analysis of the XAS lineshape.\cite{Stepanow2011} Indeed the XMCD sum rules\cite{Thole1992,Carra1993} applied to the spectra reported in \Fig{XMCD}{} reveal an orbital moment of $0.35\pm0.04$ $\mu$B and an effective spin moment of $1.03\pm0.09$ $\mu$B. We emphasize, however, that these values do not correspond to the full moment given that the paramagnetic CoTBrPP are not fully saturated in a field of 6 T. Moreover, the presence of a nonzero magnetic spin dipole moment prevents us from obtaining a precise estimate of the spin moment.\cite{Stepanow2010}

The XMCD results presented above are in contrast to those obtained with CoPc \cite{Stepanow2011} and CoTPP \cite{Vijayaraghavan2015} monolayers on noble metals, where the dichroic signal is absent. They reveal that the interaction between CoTBrPP and the Bi$_2$Te$_3$ surface is weak enough to preserve the molecular magnetic moment.

The experimental results are supported by our DFT calculations, which find a half empty $d_{z^2}$ orbital and a total spin moment of 1.11 $\mu$B. We note that the good match with the effective spin moment obtained experimentally indicates that, at this incidence angle, a finite dipole spin moment adds to the unsaturated spin moment in the XMCD spin sum rule.\cite{Thole1992,Stepanow2010} Additionally, DFT shows that the molecules have easy plane magnetic anisotropy with an energy barrier of 1.66 meV, consistent with the presence of an in-plane orbital moment (see \supp). Calculations of CoPc on Bi$_2$Te$_3$ also result in a similar spin moment of 1.09 $\mu$B, suggesting that the magnetic moment of this molecule is also preserved on this surface.

\subsection{Surface band structure}

The remaining question is then if the TSS also withstands the interfacial interactions with the magnetic molecules. In order to assess the impact of the CoTBrPP molecules on the surface electronic structure, an ARPES study was carried out as a function of molecular coverage. The latter was calculated from the attenuation of the substrate Bi 4f$_{7/2}$ peak obtained in simultaneous XPS measurements, and cross-checked by the intensity evolution of the C 1s peak (see \supp). 

\begin{figure*}
	\includegraphics{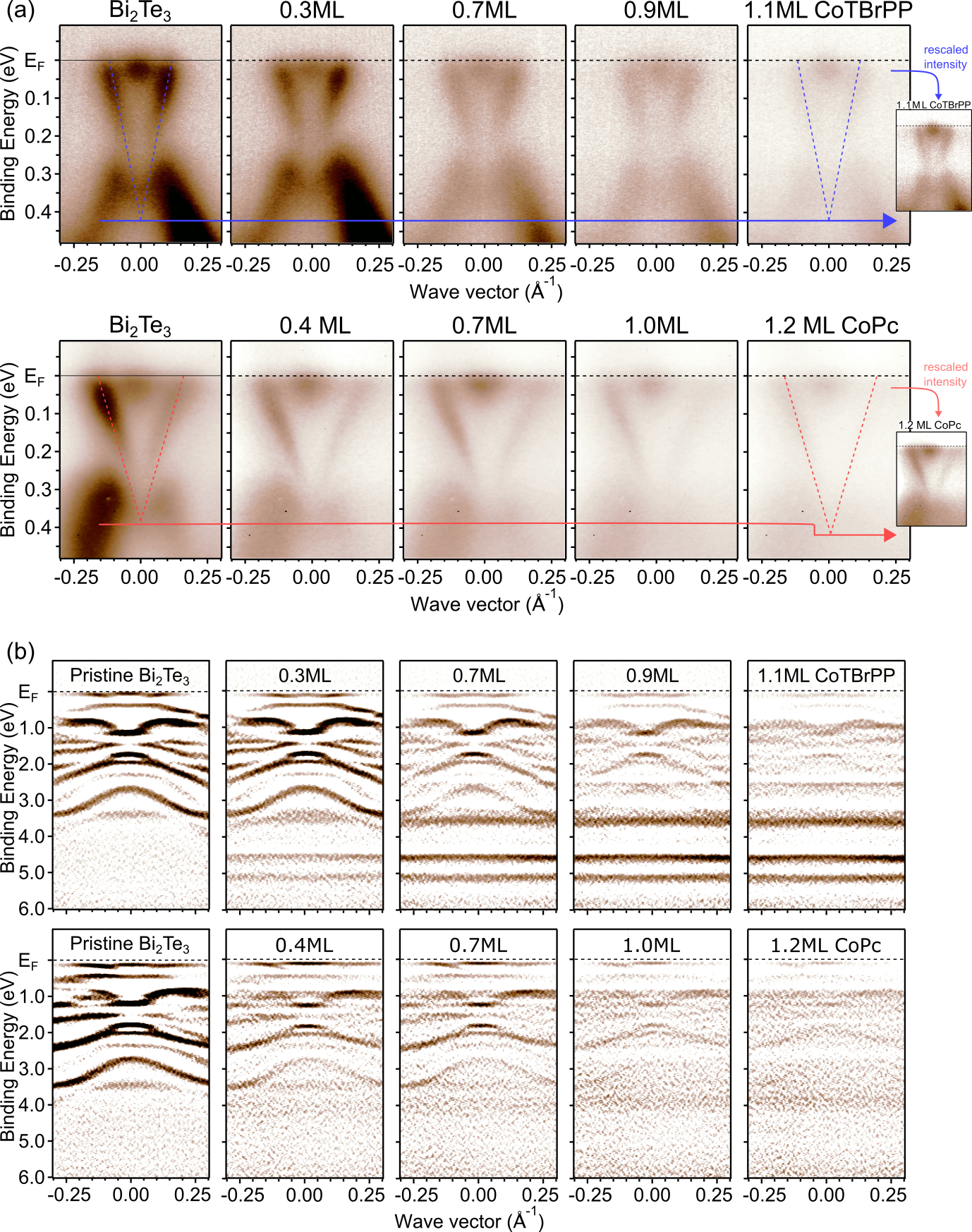}
	\caption{Coverage-dependent evolution of ARPES spectra for CoTBrPP and CoPc. (a) ARPES intensity near E$_F$ showing the TSS. (b) Second derivative of the ARPES intensity of a larger spectral range showing the evolution of Bi$_2$Te$_3$ bulk bands and molecular orbitals from the pristine surface to the monolayer completion (see \supp\ for the corresponding ARPES intensity). Blue/red dashed lines in the narrow energy range spectra of (a) are linear fits to the dispersion of the TSS Dirac cone. Corresponding horizontal lines trace the Dirac point of pristine and fully covered surfaces. Insets show spectra of the highest coverage samples with rescaled color range.}
	\label{ARPES}
\end{figure*}

\begin{figure}
	\includegraphics{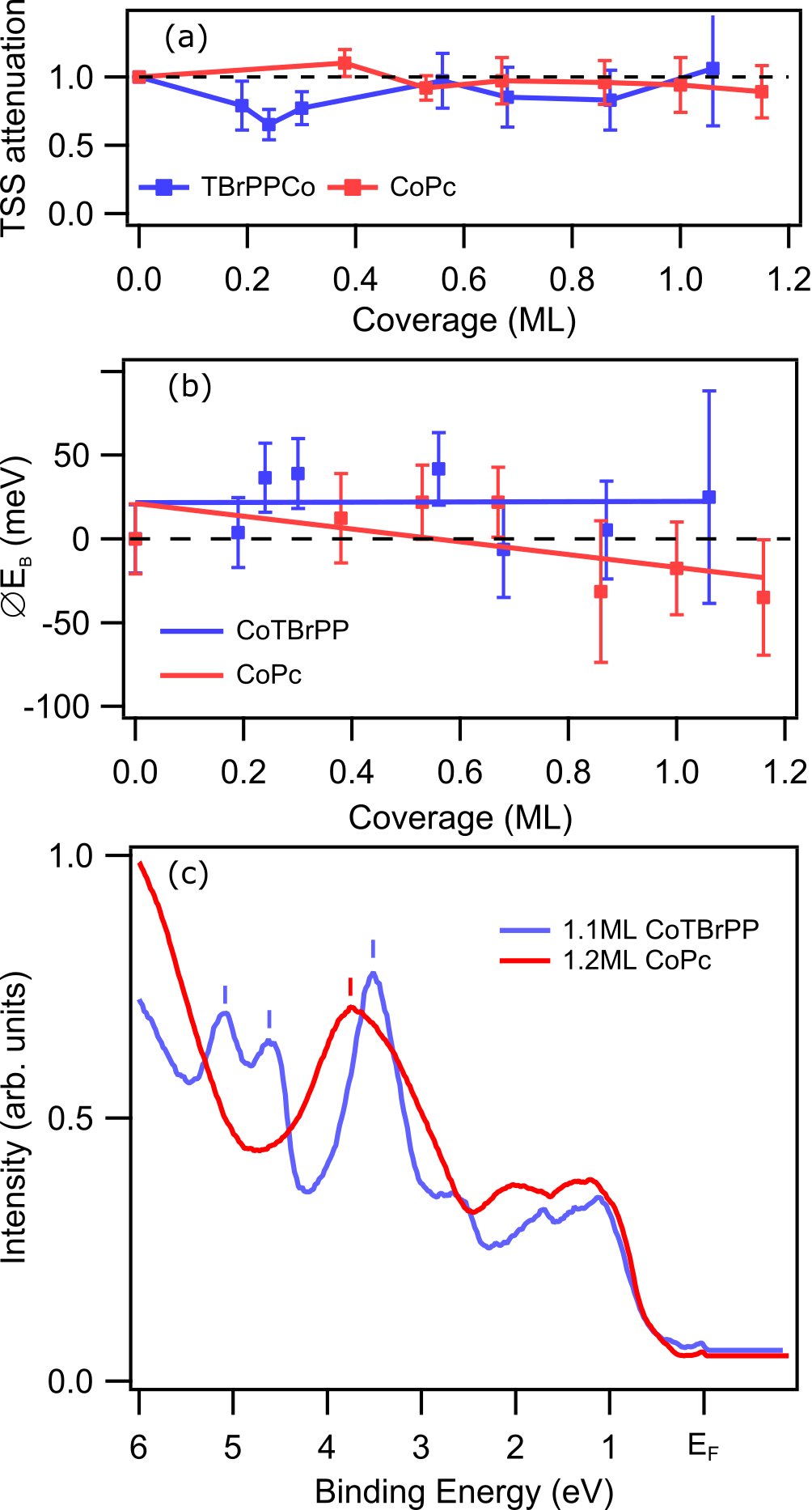}
	\caption{(a) Coverage-dependent evolution of the TSS attenuation, defined as the TSS to bulk band photoemission intensity ratio relative to that of the pristine surface. The photoemission intensity has been integrated over the areas defined in the inset ARPES map. (b) Evolution of the Dirac point obtained from the fitting of momentum distribution curves (see \supp). (c) Angle-integrated spectra of the highest coverage samples shown in \Fig{ARPES}{}. Molecular orbitals are indicated with vertical ticks.}
	\label{TSSattenuation}
\end{figure}

The characteristic Dirac cone dispersion of the Bi$_2$Te$_3$ surface state can be tracked in the narrow energy range spectra presented in \Fig{ARPES}{a}. The TSS spectral density quenches gradually with coverage. However, when normalized by the valence band intensity to account for the overall photoelectron attenuation by the molecular overlayer, the TSS intensity remains constant, as reported in \Fig{TSSattenuation}{a}. The position of the Dirac point is also unaffected by the interaction with the molecules, as can be inferred from the Dirac cones of the pristine and highest coverage samples represented by overlaid dashed lines (\Fig{ARPES}{a}). These have been determined by fitting the momentum distribution curves to Lorentzian pairs, and the resulting maxima to a linear dispersion (see \supp). As indicated by the solid arrow that connects the two Dirac points in \Fig{ARPES}{a} and the complete evolution presented in \Fig{TSSattenuation}{b}, there is no detectable charge transfer between the CoTBrPP molecular layer and the Bi$_2$Te$_3$ surface.

The lack of attenuation and charge transfer found for CoTBrPP monolayers on Bi$_2$Te$_3$ is in stark contrast to the previously reported evolution of the Bi$_2$Se$_3$ TSS upon CoPc adsorption.\cite{Caputo2016} Here the TSS is already doped at low coverage, and the spectral weight of the outermost QL that is probed by the photon energies used in our measurements is totally quenched at the monolayer completion. Motivated by such different behaviour, we extended our studies to CoPc on Bi$_2$Te$_3$. Surprisingly, the TSS presents a similar evolution as for CoTBrPP, without any evidence of attenuation relative to bulk bands (see \Fig{ARPES}{a} and \Fig{TSSattenuation}{a}). Nevertheless, we find several evidences of a stronger interaction as compared to CoTBrPP. First, the coverage dependent fit of the Dirac cone reveals a small but sizeable doping of 35 $\pm$ 28 meV at maximum coverage. A more stringent indication is provided by the molecular orbitals that are probed in the wide energy range ARPES spectra presented in \Fig{ARPES}{b}. Several molecular orbitals can be identified here as non-dispersing levels that emerge gradually with increasing coverage at a binding energy larger than 2 eV, in agreement with the onset found by STS at this energy range. Differences between the two molecular systems can be best appreciated in the angle-integrated spectra of the highest coverage samples displayed in \Fig{TSSattenuation}{c}. The broader molecular features of CoPc contrast with the sharper multiple peaks found for CoTBrPP in the same energy region, and provide a direct proof of a stronger hybridization experienced by CoPc orbitals. These conclusions are further supported by the structural differences observed in the self-assembly of the two molecular systems. Indeed, the hexagonal symmetry of the patterns found for 
CoPc \cite{Bathon2015} denotes a stronger stacking interaction than in the square CoTBrPP lattices.

The different charge transfer and hybridization found for the two molecules on Bi$_2$Te$_3$ is a direct indication of the effectiveness of ligands for fine tuning interactions. On the other hand, the different results obtained for CoPc on Bi$_2$Te$_3$ in this work, and on Bi$_2$Se$_3$ in Ref. [36] show that substrate effects can still be considerable. We attribute the stronger interactions found on Bi$_2$Se$_3$ to the larger electron affinity of Se, which favours charge transfer from the molecules to the substrate and reduces the Co-surface distance. This difference is accentuated by the different lattice constants of the two TIs, which results in a 0.15 \AA\ shorter Co-chalcogenide distance in Bi$_2$Se$_3$.

\section{Conclusions}

In summary, we find that CoTBrPP and CoPc monolayers adsorbed on Bi$_2$Te$_3$ form robust interfaces where electronic interactions can be tuned without strongly perturbing of the intrinsic properties of each constituent. Our conclusions are supported by consistent structural, electronic and magnetic information derived from a combined STM, ARPES, XMCD and DFT study. 

The observed molecular self-assembly provides a first proof that weak but finite interactions exist among the molecules and TI, which result in a close-packed square lattice arrangement of the Co ions, as preferred by the molecules, with a commensurate stacking on Te sites, as imposed by Bi$_2$Te$_3$. ARPES further reveals that the TSS remains undoped upon molecular deposition and maintains its pristine spectral intensity. The only electronic signature of interactions is the broadening of the empty d$_{z^2}$ orbital probed by STM. Interestingly, this is the orbital bearing the spin moment of the Co$^{2+}$ ion, which suggests that magnetic proximity effects may be induced by the molecules on the TI. Given that the CoTBrPP molecules are paramagnetic, however, we do not observe signatures of long-range magnetic order in our study.

%At the CoTBrPP/Bi$_2$Te$_3$ interface, the TSS remains undoped and with pristine spectral intensity. The only electronic signature of interactions is the broadening of the empty $d_{z^2}$ orbital probed by STM. Interestingly, this is the orbital bearing the persisting magnetic moment of the Co$^{2+}$ ion, suggesting that magnetic exchange can be present despite the weak interactions. The molecular self-assembly provides a final proof of weak but finite interactions, by combining the close-packed square lattice arrangement preferred by the molecules with a commensurate stacking.

The comparative ARPES study reveals stronger interactions at the CoPc/Bi$_2$Te$_3$ interface  relative to CoTBrPP/Bi$_2$Te$_3$. For CoPc, we find a significant hybridization between molecular and TI states without considerable charge redistribution. The results are consistent with the stronger stacking interaction that leads to hexagonal molecular patterns in this system.\cite{Bathon2015}

The degree of tunability offered by the planar molecular compounds spans all the way from the weak to the strong perturbation regime, as demonstrated by comparing the results presented here for Bi$_2$Te$_3$ with the case of the stronger interacting CoPc/Bi$_2$Se$_3$ interface. Finally, although intermolecular exchange interactions in this self-assembled layers are expected to be weak, we envisage that they could be enhanced by the use of planar 2D covalent or metal-organic coordination networks.

\section{Methods}

\textbf{Methods.} \textit{Sample preparation.} Twin-free Bi$_2$Te$_3$ thin films of 20 QL were synthesized by coevaporation of elemental Bi and Te (6N purity) in a molecular beam epitaxy chamber (MBE) with a base pressure of \pressure{2}{10}, equipped with thermal cracker cells.\cite{Bonell2017} Subsequently, a $20$ nm thick Te capping layer was deposited in order to prevent the TI surface oxidation during the transfer in a UHV suitcase to the different ex-situ characterization setups for the corresponding STM, ARPES and XMCD measurements. The capping layer was removed before measurements by thermal annealing at $130^{\circ}$C (see \supp). CoTBrPP molecules were synthesized in-house (see \supp) and CoPc were commercial (Sigma-Aldrich). Both were deposited on the freshly decapped TI thin films at a base pressure of \pressure{5}{10}, with the substrate kept at 77 K and subsequently annealed to room temperature to promote the self-assembly.
\\ \textit{ARPES and XPS measurements} were performed with a Phoibos 150 analyzer (SPECS GmbH, Berlin, Germany) at a temperature and base pressure of $77$ K and \pressure{5}{10}, with monochromatic HeI UV ($h\nu=21.2$ eV) and Al $K_{\alpha}$ ($h\nu=1486.6$ eV) sources at incidence angles of $55^{\circ}$ and $50^{\circ}$ respectively. Measurements were performed along the $\Gamma-K$ direction for CoTBrPP films, and the $\Gamma-M$ direction for CoPc films. 
\\ \textit{STM measurements} were carried out at a temperature and base pressure of $5.4$ K  and \pressure{5}{10}. $dI/dV$ spectra and maps were acquired at constant height mode, using a lock-in at a modulation voltage of $V_{mod}$=20 mV. 
\\ \textit{XAS and XMCD measurements} were carried out at the BOREAS beamline of the Alba Synchrotron facility.\cite{Barla2016} Spectra were taken in total-electron-yield mode with right ($I^+$) and left ($I^-$) circularly polarized photons at normal ($0^{\circ}$) and grazing ($70^{\circ}$) incidence, in the presence of a magnetic field up to $\pm 6$ T aligned parallel to the incident beam, and at a temperature and base pressure of $7.5$ K and \pressure{1}{10}. XAS and XMCD spectra are represented as the average and difference of the negative and positive circularly polarized absorption, respectively. A molecular coverage of $\lesssim 1$ ML was estimated from the XAS $L_3$ edge jump, previously calibrated with in-situ STM measurements of the same molecules on Au(111).
\\ \textit{Ab initio calculations} were performed in the framework of the density functional theory (DFT) as implemented in VASP.\cite{kresse_efficient_1996} Core electrons were treated using the projector augmented-wave method.\cite{blochl_projector_1994,kresse_ultrasoft_1999} For the exchange and correlation functional the PBE flavor of the generalized gradient approximation\cite{perdew_generalized_1996} was used. The Dudarev form of the GGA+U method\cite{dudarev_electron-energy-loss_1998} was applied to improve the description of the \textit{3d} electrons of cobalt with U-J=3~eV. Missing van der Waals forces were included using the Tkatchenko-Scheffler method.\cite{tkatchenko_accurate_2009} Wave functions were expanded using a plane wave basis set with an energy cutoff of 400~eV. The Brillouin zone was sampled with a ($3 \times 3 \times 1$) k-point mesh. The geometrical structures were relaxed until forces were smaller than 0.01~eV/\AA. Magnetic anisotropy energies were determined by total energy differences after including spin-orbit coupling in the calculations,\cite{steiner_calculation_2016} and using a ($6 \times 6 \times 1$) k-point mesh. Magnetic moments were determined by performing a Bader analysis.\cite{tang_grid-based_2009} Drawings of the structures were produced by VESTA.\cite{Momma:db5098}

\acknowledgement

The authors thank G. Ceballos and M. Maymó for their support in the experimental development. This research was funded by the CERCA Program/Generalitat de Catalunya, and supported by the Spanish Ministry of Economy and Competitiveness, MINECO (under Contracts No. MAT2016-78293-C6-2-R, MAT2016-75952-R, RTI2018-095622-B-I00, and Severo Ochoa No. SEV-2017-0706), the Secretariat for Universities and Research, Knowledge Department of the Generalitat de Catalunya 2017 SGR 827, the European Regional Development Fund (ERDF) under the program Interreg V-A Espa\~{n}a-Francia-Andorra (Contract No. EFA 194/16 TNSI), the European Commission under the H2020 FET Open project \textit{Mechanics with Molecules} (MEMO) (Grant No. 766864), TOCHA (Grant No. 824140), and the European Research Council under (Grant No. 306652 SPINBOUND).

\suppinfo

Decapping of the Te protective layer, molecular coverage calibration, fit of Dirac cone dispersion, extended DFT PDOS, adsorption configurations and magnetic anisotropy data, synthesis of CoTBrPP. 
\bibliography{manuscript4BrCoTPPonBi2Te3}

\end{document}